# Three-dimensional surface topography of graphene by divergent beam electron diffraction

Tatiana Latychevskaia[1,*], Wei-Hao Hsu[2,3], Wei-Tse Chang[2], Chun-Yueh Lin[2] and Ing-Shouh Hwang[2,3,*]

[1]Physics Department of the University of Zurich, Winterthurerstrasse 190, CH-8057 Zürich, Switzerland

[2]Institute of Physics, Academia Sinica, Nankang, Taipei 115, Taiwan

[3]Department of Materials Science and Engineering, National Tsing Hua University, Hsinchu 300, Taiwan

*Corresponding authors: tatiana@physik.uzh.ch, ishwang@phys.sinica.edu.tw

**ABSTRACT**

There is only a handful of scanning techniques that can provide surface topography at nanometre resolution. At the same time, there are no methods that are capable of non-invasive imaging of the three-dimensional surface topography of a thin free-standing crystalline material. Here we propose a new technique - the divergent beam electron diffraction (DBED) and show that it can directly image the inhomogeneity in the atomic positions in a crystal. Such inhomogeneities are directly transformed into the intensity contrast in the first order diffraction spots of DBED patterns and the intensity contrast linearly depends on the wavelength of the employed probing electrons. Three-dimensional displacement of atoms as small as 1 angstrom can be detected when imaged with low-energy electrons (50 – 250 eV). The main advantage of DBED is that it allows visualisation of the three-dimensional surface topography and strain distribution at the nanometre scale in non-scanning mode, from a single shot diffraction experiment.

**INTRODUCTION**

Measurements of surface topography with atomic resolution is absolutely crucial for many branches of science, including physics, chemistry and biology. Free-standing graphene, with its intrinsic and extrinsic ripples, offers an ideal test object for any three-dimensional surface mapping technique. According to the Mermin-Wagner theorem, at any finite temperature two-dimensional materials must exhibit intrinsic corrugations[1]. Such intrinsic ripples with a period of about 5–10 nm were predicted for free-standing graphene by Monte Carlo simulations[2]. In 2007, Meyer *et al.* performed convergent beam electron diffraction (CBED) imaging of graphene, observing intensity variations in the first-order diffraction spots[3-4]. CBED was realised in a conventional TEM where the electron beam spatial coherence was about 10 nm. These intensity variations were explained by changes in the local orientation of graphene, namely by its deflection within ±2° from the normal to the flat surface[3] or by 0.1 rad[4], which could be attributed to ripples with the amplitude of 1 nm at 20 nm length.

There are a number of techniques that allow surface topography to be measured with nanometre resolution, such as scanning tunnelling microscopy (STM)[5] and atomic force microscopy[6]. Furthermore, some of them have been applied to image ripples in graphene[7]. However, it is not possible to call such techniques completely non-invasive. Graphene has very low bending rigidity, and the probe can affect the ripple distribution. This is, for instance, the case in STM imaging of free-standing graphene[8], where strong interaction between the STM tip and the graphene can even lead to the flipping of ripples[9]. Also, scanning techniques, because of slow scanning speed, intrinsically incapable of obtaining temporal dynamics across the studied surface, which is expected for flexural phonon modes[10]. So far, only the temporal dynamics of a ripple at the fixed tip position[11] has been obtained by STM. Thus, there is no technique that allows the three-dimensional surface topography of a thin free-standing crystalline material to be detected in a non-invasive and non-scanning mode.

We propose divergent beam electron diffraction (DBED), which can be realised for thin crystalline samples in transmission mode. DBED allows imaging of a few hundreds nm$^2$ area in a non-invasive and non-scanning mode, such as a single-shot diffraction experiment, and can be applied for the observation of the temporal dynamics of the three-dimensional surface.

**RESULTS**

*Condition for observation of divergent beam electron diffraction (DBED) patterns*

To understand the formation of the intensity contrast in DBED we consider diffraction of electron wavefront by a periodic lattice of graphene. When a plane wave scatters off a periodic sample, diffraction peaks are observed in the far-field intensity distribution. Conventionally, the distribution of the intensity in a diffraction pattern is presented in **k**- coordinates, where $k = \frac{2\pi}{\lambda}\sin\vartheta$, $\lambda$ is the wavelength and $\vartheta$ is the scattering angle. The positions of the diffraction peaks are determined by fulfilling the Bragg condition or alternatively by the Ewald's sphere construction. For a periodic lattice with period $a$, the corresponding reciprocal lattice points are found at $k = 2\pi/a$ in the reciprocal space. Whether the corresponding diffraction peaks are observed on a detector or not is determined by the *k*-component range of the imaging system $k_{max} = \frac{2\pi}{\lambda}\sin\vartheta_{max}$. For example, for graphene, the six first-order diffraction peaks at $k_1 = 2\pi/a_1$ are associated with diffraction at crystallographic planes with the period $a_1 = 2.13$ Å. To detect these first-order diffraction peaks of graphene, the wavelength of the imaging electrons ($\lambda$), and the maximum acceptance angle of the imaging system ($\vartheta_{max}$) must be selected such that $k_{max} > k_1$:

$$\frac{\sin\vartheta_{max}}{\lambda} > \frac{1}{a_1}. \tag{1}$$

When the incident wave is not a plane wave, but rather diverges, as in the DBED regime, each diffraction peak turns into a finite-size intensity spot, but the positions of the spots remain the same as the positions of the diffraction peaks, as illustrated in Fig. 1a. The intensity distribution within one DBED spot reflects the deviation of the atom distribution from the perfect periodic positions.

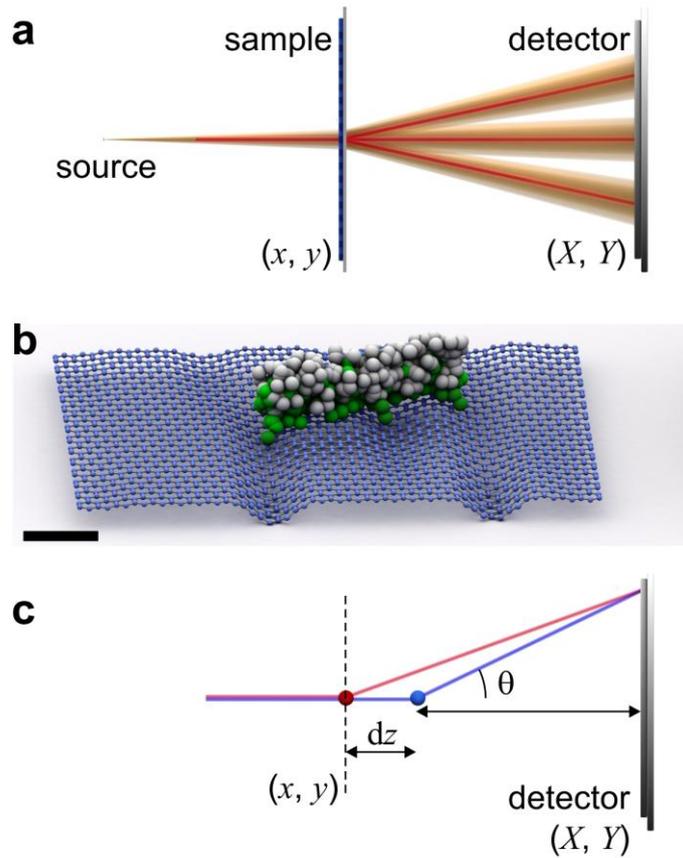

**Figure 1.** Principle of divergent beam electron diffraction (DBED) imaging. **a**, Illustration of beam propagation in diffraction mode (red) and in DBED mode (orange). **b**, Representation of an adsorbate on graphene causing strain and ripples. **c**, Geometrical arrangement of scattering from two atoms positioned at different $z$-distances. The scale bar in **b** corresponds to 1 nm.

If graphene is rippled, the carbon atoms deviate from their perfect lattice positions in all three dimensions. Such ripples can be intrinsic, or be caused by, for instance, an adsorbate on the surface of graphene, Fig. 1b. Two waves scattered off two atoms at different $z$-positions travel across different optical paths to a certain point on a detector, as illustrated in Fig. 1c. The difference in the optical paths amounts to

$$\Delta s = \mathrm{d}z\left(1-\cos\theta\right), \qquad (2)$$

where $\mathrm{d}z$ is the difference in $z$-positions of the atoms, and $\theta$ is the scattering angle. The intensity of the formed interference pattern is proportional to the relative phase shifts between the scattered waves $\Delta\varphi = k\Delta s$. For small scattering angles, as for example in the zero-order diffraction spot, the phase shift is negligible and no intensity contrast variations are expected. For higher scattering angles, as in the first-order DBED spot, the phase shift becomes significant, thus leading to noticeable intensity variations. Therefore, the intensity in the zero-order DBED spot is not sensitive to variations in the $z$-positions of the atoms, whereas the intensity distribution in the higher-order DBED spots is highly sensitive to the distribution of atomic *z*-positions, see also the Supplementary Fig. 1. This holds for any wavelength of the imaging electrons. Although we present experimental data acquired with low-energy electrons (230 and 360 eV), similar DBED patterns can be acquired with high-energy electrons in a conventional TEM. It must be pointed out that lower-energy electrons are more sensitive to the distribution of *z*-positions of atoms in a two-dimensional material. For example, for graphene, a ripple of height $h$ will cause intensity variations in the first-order DBED spot because of a superposition of the scattered waves with the phase shift

$$\Delta\varphi = kh(1-\cos\theta) = \frac{2\pi}{\lambda}h\left[1-\sqrt{1-\left(\frac{\lambda}{a_1}\right)^2}\right] \approx \frac{\pi}{a_1^2}h\lambda. \qquad (3)$$

Thus the phase shift, and therefore the contrast of the formed interference pattern, depends on the wavelength of the probing wave and decreases as the energy of the probing waves increases. For low-energy electrons, even a small height $h$ of the ripples will cause a significant phase shift and noticeable interference pattern in a first-order DBED spot. For example, a ripple with $h = 1$ Å, when imaged with electrons of 230 eV kinetic energy, will cause a phase shift of about $\Delta\varphi = 0.5$ radian in the first-order DBED spot. Another advantage of using low-energy electrons is their relatively low radiation damage.[12]

*Experimental realization of divergent beam electron diffraction (DBED)*

Recently, it was reported that electron point projection microscopy (PPM), which is also a Gabor-type in-line holography[13-17], has been applied to image graphene. In certain experimental geometrical arrangements PPM resulted in a very bright central spot and six first-order diffraction spots[18]. However, no quantitative explanation for intensity variations within the diffraction spots was provided. Figure 2a shows the experimental arrangement of the low-energy electron point projection microscope used in this work, which has been described elsewhere[18]. The electron beam is field emitted from a single-atom tip (SAT)[19-20], in this case we used an iridium-covered W(111) SAT[21,22]. This type of SAT has been demonstrated to provide high brightness and fully spatially coherent electron beams[21,22] with Gaussian distributed intensity profiles and a full divergence angle of 2–6°. We studied free-standing monolayer graphene stretched over a hole in a gold-coated $Si_3N_4$ membrane, (for preparation procedure see[18]). When the tip is positioned at a short distance (microns or smaller) in front of the sample, the transmitted electron beam forms a magnified projection image of the illuminated sample (zero-order spot pattern) at the detector, as shown in Fig. 2b. The magnification is given by $M = D/d$, where $d$ is the source-to-sample distance and $D$ is the source-to-detector distance. A DBED pattern is observed when parameters of the experimental setup satisfy Eq. (1). The size of the zero- and first-order DBED spots is given by the size of the imaged area (limited either by the size of the probing beam or by the size of the sample supporting aperture) multiplied by the magnification $M$.

Figure 2 shows DBED pattern recorded at $t = 0$, 200 and 500 s. Fig. 2c presents the central spot recorded at $t = 0$, 200 and 500 s and Fig. 2d shows the sample distribution obtained by reconstruction of the central spot at $t = 0$ s. The dark distributions on the right and left edges in the zero-order DBED spots can be associated with the aggregation of adsorbates on graphene, which are non-transparent for the electron beam. The centre region in the zero-order DBED spot is formed by the electron wave transmitted mainly through a clean graphene region with only one or two darker or brighter spots corresponding to small individual adsorbates[23]. The dark distributions associated with adsorbate aggregates remain visible in the first-order DBED spots. However, in addition, bright and dark stripes are evident in the region between the dark distributions. As demonstrated above, the first-order DBED spots (Fig. 2e–g) exhibit intensity contrast variations that are not observed in the corresponding area of the zero-order DBED spot, which agrees well with the aforementioned explanations that waves scattered off atoms positioned at different $z$-positions, contribute to the contrast formation at high scattering angles. Also, it should be noted that the first-order DBED spots have slightly different intensity distributions between themselves. From the data presented in Fig. 2e – g, it can be seen that the intensity distribution within the first-order DBED spots also varies in time, probably due to changes in the distribution of the adsorbates.

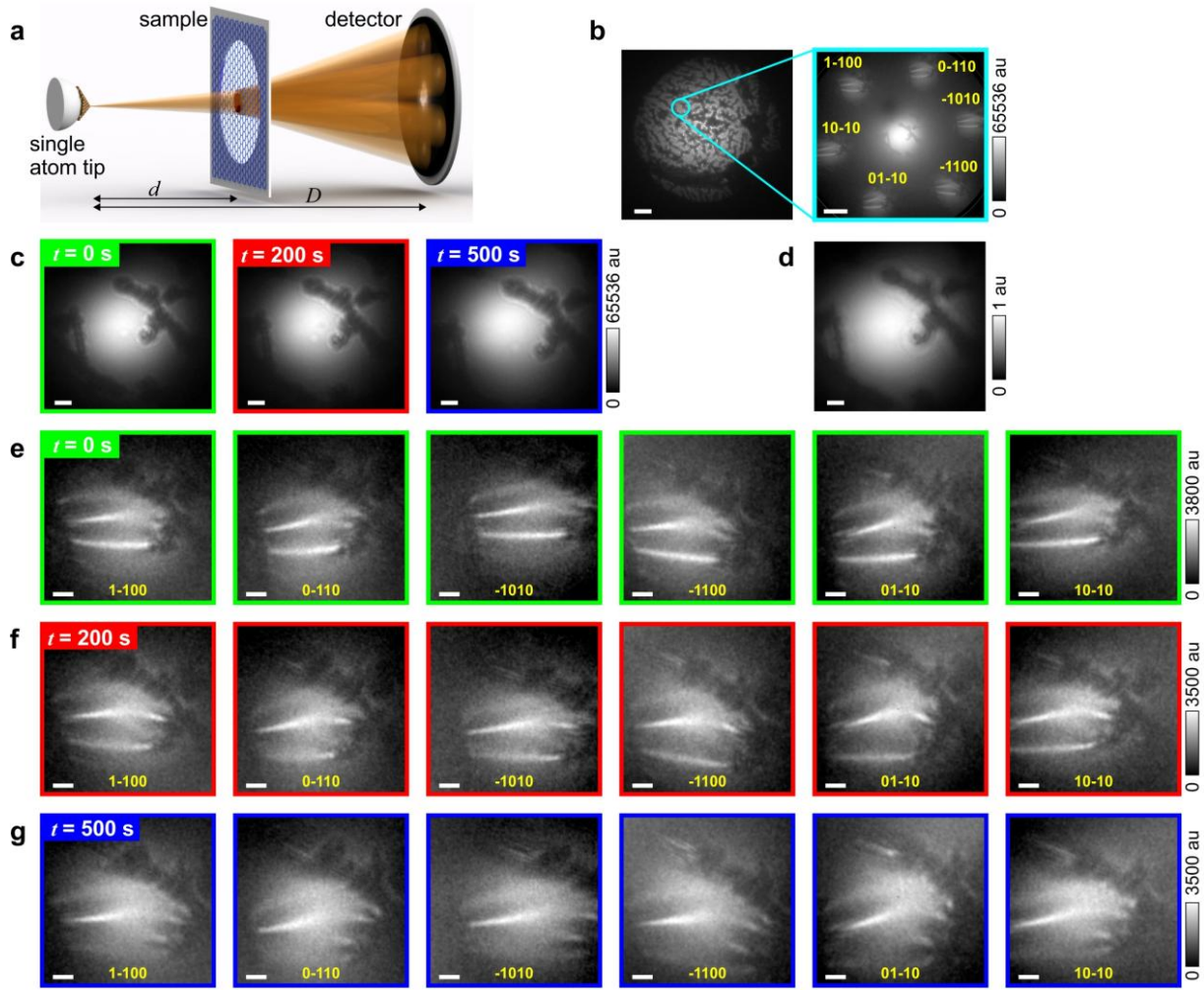

**Figure 2.** Divergent beam electron diffraction (DBED) patterns of graphene with low-energy electrons. **a**, Experimental scheme for low-energy lens-less coherent electron microscopy, which comprises a single-atom tip, graphene sample and detector; further details are provided in the Methods. **b**, Point-projection microscopy (PPM) image recorded at low magnification when the tip is far from the sample (left) and the DBED pattern of a selected region after the tip is moved close to the sample (right), so that Eq. (1) is fulfilled. The PPM image is recorded with electrons of 360 eV and the source-to-detector distance is 142 mm, the scale bar corresponds to 200 nm. The DBED pattern is recorded with electrons of 230 eV and the source-to-detector distance is 51 mm, the scale bar corresponds to 100 nm. The DBED pattern is shown with a logarithmic intensity scale because the intensity of the zero-order diffraction spot is about 50 times greater than that of the first-order diffraction spots. **c**, The zero-order DBED spots recorded at $t = 0$, 200 and 500 s and **d**, a reconstruction of the central region of the DBED pattern recorded at $t = 0$ sat a source-to-sample distance of about 550 nm obtained numerically by an algorithm explained elsewhere[24]. The size of the illuminated area approximately corresponds to the size of the shown reconstruction, $168 \times 168$ nm$^2$. **e** – **g**, The first-order DBED spots recorded at $t = 0$, 200 and 500 s, respectively. The scale bars in **c** – **g** correspond to 20 nm.

*Simulated divergent beam electron diffraction (DBED) patterns*

To characterise the observed ripples quantitatively, we performed numerical simulations of DBED patterns. The diffracted wavefront at the detector is simulated using the following distribution:

$$U(X,Y) = \frac{i}{\lambda} \iint t(x,y,z) \frac{\exp\left(\frac{2\pi i}{\lambda}\sqrt{x^2+y^2+z^2}\right)}{\sqrt{x^2+y^2+z^2}} \frac{\exp\left(\frac{2\pi i}{\lambda}\sqrt{(X-x)^2+(Y-y)^2+(Z-z)^2}\right)}{\sqrt{(X-x)^2+(Y-y)^2+(Z-z)^2}} \mathrm{d}x\mathrm{d}y,$$

(4)

where $t(x,y,z)$ is the transmission function in the object domain, $(x,y,z)$ are the coordinates in the sample domain and $(X,Y)$ are the coordinates in the detector plane. $t(x,y,z)$ includes the aperture distribution $A(x,y)$ and the carbon atom distribution in graphene $G_0(x_i, y_i, z_i)$, where $G_0(x_i, y_i, z_i)$ is 1 at the position $(x_i, y_i, z_i)$ of carbon atom $i$ and 0 elsewhere. Details of the simulation are provided in the Methods. Note that the simulations were performed assuming a monochromatic spatially coherent electron source. No other approximations and no fast Fourier transforms are used in this simulation.

The following device configurations were simulated and are shown in Figs. 3 and 4: clean graphene (Fig. 3a–b), graphene with a single adsorbate (Fig. 3c–f), graphene with an out-of-plane ripple (Fig. 4a–f) and graphene with an in-plane ripple (Fig. 4g–i). In all simulated DBED patterns, the intensity of the zero-order DBED spot is about 50 times higher than that of the first-order DBED spots, which agrees with the experimental observations. The first-order DBED spots appear to be distorted because of the geometrical conditions selected in the simulations: plane detector and relatively short distance between the source and the sample.

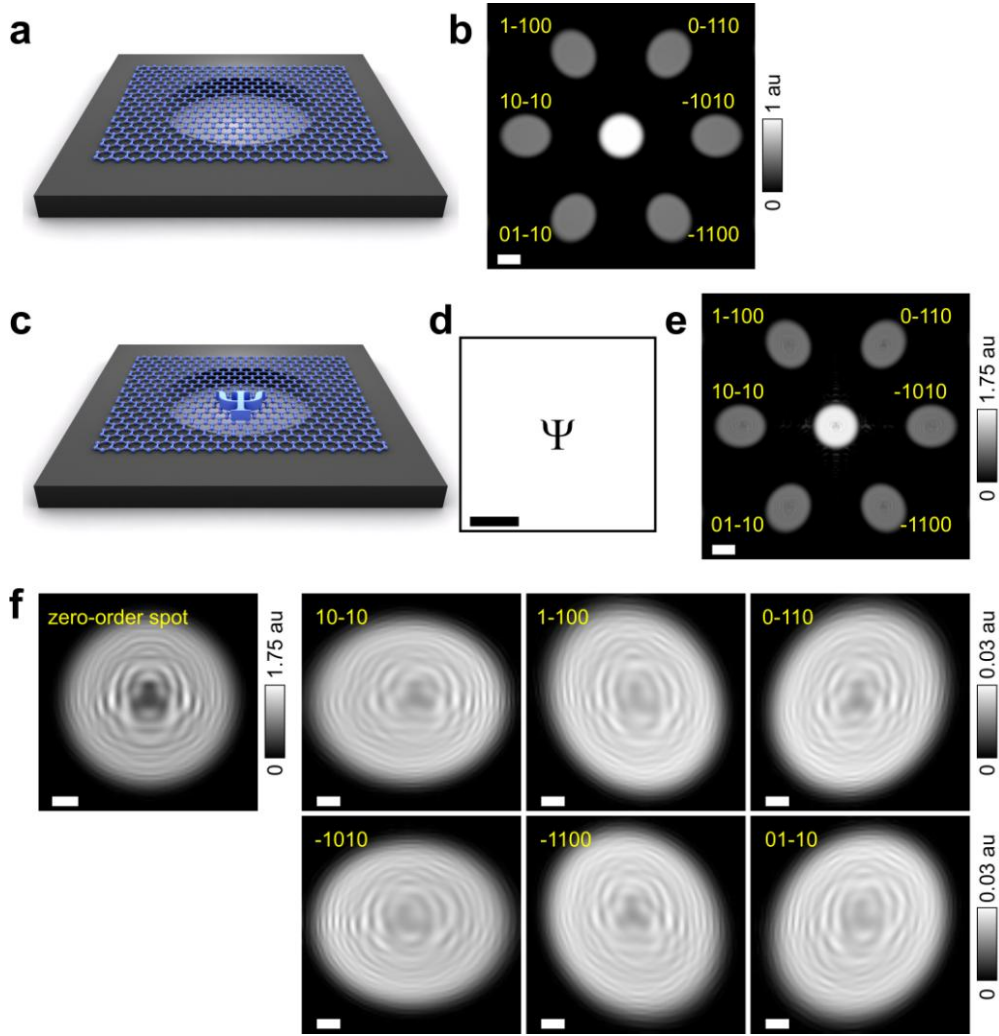

**Figure 3.** Simulated divergent beam electron diffraction (DBED) patterns of graphene with an adsorbate. **a**, Sketch of graphene stretched over an aperture. **b**, Full DBED pattern of graphene stretched over an aperture shown in logarithmic intensity scale. **c**, Sketch of graphene sample with an adsorbate in the form of the letter psi. **d**, Distribution of the transmission function of the adsorbate. **e**, Full DBED pattern of graphene with an adsorbate in the form of the letter psi shown in logarithmic intensity scale. **f**, Magnified zero-order (upper left) and first-order DBED spots of the DBED pattern shown in **e**. The Miller indices indicate the diffraction spots. In the simulations, the aperture diameter is 40 nm, the source-to-sample distance is 200 nm, the source-to-detector distance is 70 mm, and the electron energy is 230 eV. The simulations are done using Eq. (4), the details of the simulations are provided in the Methods. The scale bars in **b** and **e** correspond to 20 nm, the scale bar in **d** corresponds to 10 nm and the scale bars in **f** correspond to 5 nm.

A simulated DBED pattern of perfectly planar clean graphene stretched over an aperture, as illustrated in Fig. 3a, is shown in Fig. 3b. A simulated DBED pattern of graphene with an opaque adsorbate in the form of the letter $\Psi$ on its surface (as illustrated in Fig. 3c–d), is shown in Fig. 3e. From Fig. 3e–f, it can be seen that the presence of an adsorbate changes the contrast of the zero-order DBED spot, which appears as an in-line hologram of the adsorbate.

The simulated first-order DBED spots exhibit similar contrast variations as the zero-order diffraction spot, though less pronounced. Thus, the presence of an adsorbate creates almost the same intensity distribution in all DBED spots.

In the real experimental situation an adsorbate on graphene can cause a certain amount of strain and hence additional rippling in graphene. We simulated two kinds of ripples: out-of-plane and in-plane. In the simulations, the carbon atoms are shifted from their perfect lattice positions $G_0(x_i, y_i, z_i)$ in the graphene plane. Each ripple is described by a Gaussian distribution with amplitude $h$ and standard deviation $\sigma$. The results of the simulations are shown in Fig. 4.

Two simulated out-of-plane ripples are sketched in Figs. 4a and 4d: in the negative $z$-direction (towards the electron source) and in the positive $z$-direction (away from the electron source). The parameters of the ripples are $h = 3$ Å and $\sigma = 1.5$ nm. The corresponding simulated DBED patterns are shown in Figs. 4b and 4e, and the magnified first-order DBED spots are shown in Figs. 4c and 4f, respectively. From Figs. 4b and 4e it can be seen that there is no intensity variations within the zero-order DBED spot, which is similar to the behaviour of the zero-order DBED spot intensity of a clean graphene region in experimental and simulated DBED patterns: Figs. 2c and 3b, respectively. At the same time, all the first-order DBED spots demonstrate very similar intensity variations that resemble the original ripple distribution. Such resemblance can be explained by considering the ripple as a phase-shifting object that changes the phase of the incident wave by $\Delta\varphi(x, y)$. It has been shown that the phase change introduced to the incident wave in the object domain is preserved while the wave propagates toward the detector[25]. Within the approximation of a weak phase-shifting object $e^{i\Delta\varphi(x,y)} \approx 1 + i\Delta\varphi(x, y)$. The introduced phase change is transformed into intensity contrast at the detector as $|\Delta\varphi(X,Y)|^2$, as explained in the Supplementary Note 1. For a ripple of height $h$, the expected intensity distribution is given by $\Delta I(X,Y) \approx |\Delta\varphi(X,Y)|^2 \approx \left(\frac{2\pi}{\lambda} h(1 - \cos\vartheta)\right)^2$. From these simulations, one can conclude that the three-dimensional form of a ripple can be directly visualised from the intensity distribution in the first-order DBED spot: a ripple towards the source results in a darker intensity distribution (Fig. 4a–c), while a ripple towards the detector results in a brighter intensity distribution (Fig. 4d–f). For example, most of the ripples in the experimental distributions shown in Fig. 2 are out-of-plane ripples towards the detector. The intensity profiles of the DBED patterns of the out-of-plane ripples of amplitude 1 and 2 Å are provided in the Supplementary Fig. 2, demonstrating that ripples of 1 Å are already sufficiently strong to cause noticeable intensity variations in the DBED pattern.

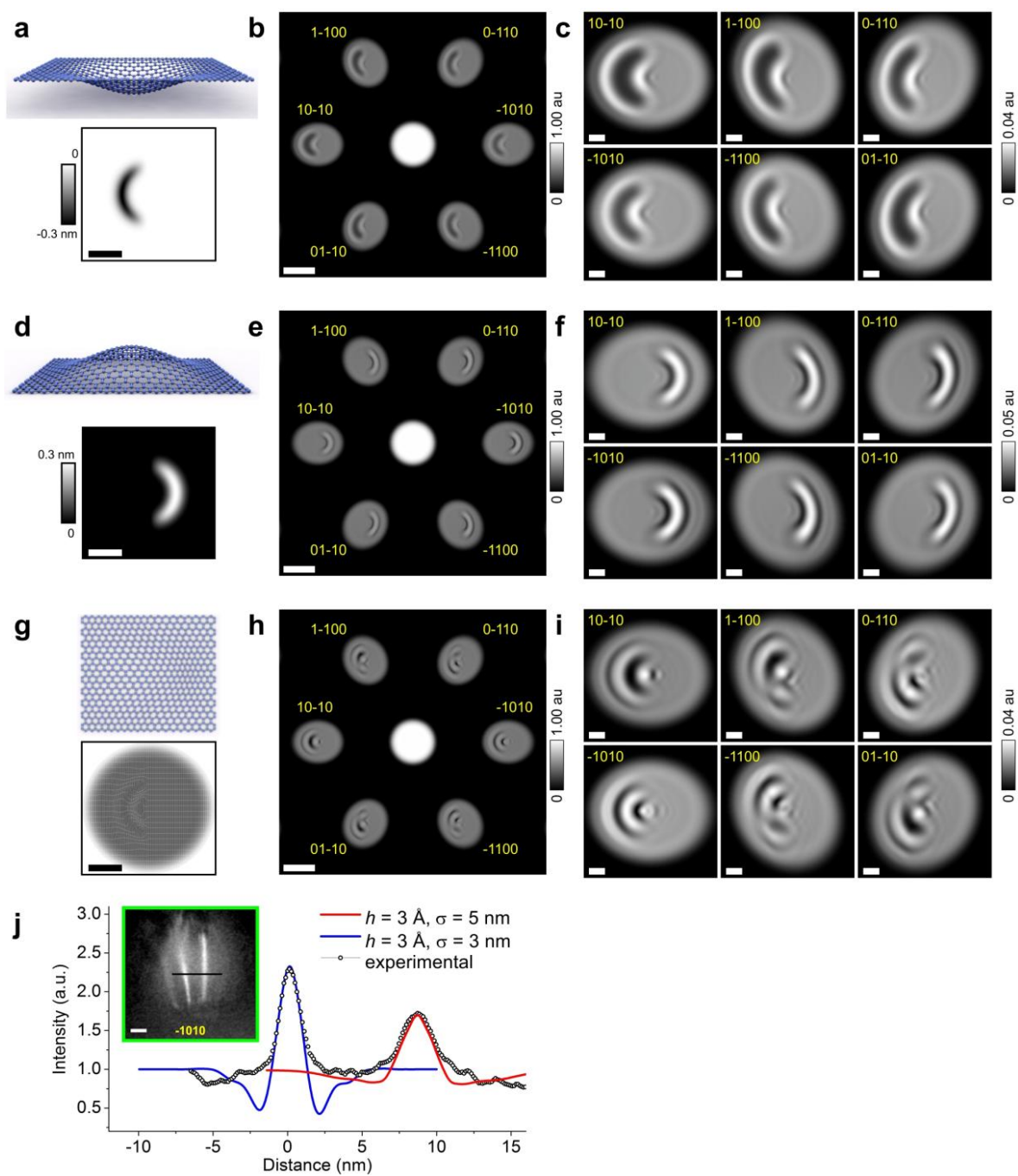

**Figure 4.** Simulated divergent beam electron diffraction (DBED) patterns of graphene with ripples. **a**, A negative out-of-plane ripple (towards the electron source) in graphene: sketch of atomic displacements and distribution of atomic $z$-shifts. **b**, Simulated full DBED pattern of graphene with a negative out-of-plane ripple shown with a logarithmic intensity scale, and **c**, magnified first-order diffraction spots. **d**, A positive out-of-plane ripple in graphene: sketch of atomic displacements and distribution of atomic $z$-shifts. **e**, Simulated full DBED pattern of graphene with a positive out-of-plane ripple shown with a logarithmic intensity scale, and **f** magnified first-order DBED spots. **g**, An in-plane ripple (in the $(x, y)$-

plane): sketch of atomic displacements and top view of the atomic distribution of Carbon atoms used in the simulations. **h**, Simulated full DBED pattern of graphene with an in-plane ripple shown with a logarithmic intensity scale, and **i** magnified first-order DBED spots. In the simulations shown in **b**, **c**, **e**, **f**, **h** and **i**, the aperture diameter is 40 nm, the source-to-sample distance is 200 nm, the source-to-detector distance is 50 mm, and the electron energy is 230 eV. The Miller indices indicate the diffraction spots. **j**, Intensity profile through the centre of the (-1010) spot of the experimental DBED pattern shown in Fig. 2e fitted with profiles of two simulated DBED patterns of two ripples ($h = 3$ Å, $\sigma = 5$ nm and $h = 3$ Å and $\sigma = 3$ nm); in these simulations, the source-to-sample distance is 550 nm, the source-to-detector distance is 51 mm, and the electron energy is 230 eV. The simulations are done using Eq. (4), the details of the simulations are provided in the Methods. The scale bars in **a**, **d**, and **g** correspond to 10 nm, the scale bars in **b, e, h** and **j** correspond to 20 nm and the scale bars in **c, f** and **i** correspond to 5 nm.

In the simulation of an in-plane ripple, sketched in Fig. 4g, the positions of carbon atoms are shifted laterally in the $(x, y)$-planes in the form of a Gaussian-distributed profile with parameters $h = 1$ Å and $\sigma = 5$ nm. The corresponding DBED pattern is shown in Fig. 4h, and the magnified first-order DBED spots are shown in Fig. 4i. Also, for this type of ripple the intensity of the zero-order DBED spot does not show any variations, similar to the experimental and simulated DBED patterns shown in Fig. 2c and Fig. 3b, respectively. The first-order DBED spots show pronounced intensity variations, which only slightly resemble the original ripple distributions. Unlike in the case of the out-of-plane ripples, here the intensity distributions within different first-order DBED spots are very different from each other. A similar effect is observed in the experimental images. This means that in-plane ripples also contribute to the contrast formation in the first-order DBED spots in the experimental images.

An intensity profile through one of the first-order DBED spots in the experimental data shown in Fig. 2e was fitted with two profiles of simulated DBED patterns of two out-of-plane ripples. Good matching between the distributions of the simulated and experimentally acquired intensity peaks was observed when the ripples were set to have height $h = 3$ Å and standard deviation of $\sigma = 5$ nm and $\sigma = 3$ nm; see Fig. 4j.

Experimental parameters such as distances and energies influence the intensity contrast on the detector. The same ripple can produce different intensity contrast on a detector when, for example, the source-to-sample distance is varied; see the additional simulations provided in the Supplementary Fig. 2.

**DISCUSSION**

To summarize, we showed that in DBED patterns the contribution from the adsorbates and three-dimensional distribution of ripples in graphene can be separated by comparing the intensity in the zero- and the first-order DBED spots. DBED imaging allows direct visualisation of the distribution of the ripples in graphene, and thus may provide accurate mapping of the three-dimensional topography of free-standing graphene. The intensity contrast within the first-order DBED spots linearly depends on the wavelength of the probing electrons. Thus, low-energy electrons are highly sensitive to inhomogeneities in the spatial distribution of atoms. When imaged with low-energy electrons, the ripples of amplitude 1 Å are sufficiently strong to cause noticeable intensity variations in the first-order diffraction spots of a DBED pattern. Thus, very weak ripples associated with strain caused by adsorbates on the graphene surface can be directly visualised and studied by DBED. In the experimental images, we observe ripples mainly formed between adsorbates, as though the adsorbates wrinkle the graphene surface around themselves. We also experimentally observed that the distribution of ripples, and, hence, the strain distribution, varies over time.

The out-of-plane and in-plane ripples are distinguishable by their appearence in the DBED patterns. Out-of-plane ripples produce similar intensity distributions between all the first-order DBED spots, whereas the in-plane ripples produce different intensity distributions for different first-order DBED spots. From comparison of the experimental results with the simulations, we conclude that the graphene surface is deformed mainly by out-of-plane ripples with a small contribution from the in-plane ripples. From the intensity of the ripple (darker or brighter), its direction can be estimated. In both simulations and experiments, the six first-order DBED spots exhibit slightly different distributions of intensity, which suggests that each DBED spot carries information about the three-dimensional surface illuminated at a slightly different angle. Thus, it should be possible to reconstruct a three-dimensional distribution of graphene surface directly from its DBED pattern. This means that DBED provides a unique tool that allows the three-dimensional topography and thus the three-dimensional strain distribution to be studied at the nanometre scale. Although we have demonstrated DBED with low-energy electrons, DBED can be also realised with high-energy electrons in conventional TEM setups.

## METHODS

**Low-energy DBED experimental arrangement.** The microscope is housed in an ultra-high vacuum (UHV) chamber. A single-atom tip (SAT) is mounted on a three-axis piezo-driven positioner (Unisoku, Japan) with a 5 mm travelling range in each direction. The detector consists of a micro-channel plate (Hamamatsu F2226-24PGFX, diameter = 77 mm) and a phosphorous screen assembly. The detector is mounted on a rail and can be moved along the beam direction. A camera (Andor Neo 5.5 sCMOS, 16-bit, 2560 × 2160 pixels) adapted with a camera head (Nikon AF Micro-Nikkor 60 mm f/2.8 D) is placed behind the screen outside the UHV chamber to record the images on the screen. The whole system was kept at room temperature during electron emission, and the base pressure of the chamber is around $1\times10^{-10}$ Torr.

**Simulation procedure.** $L(...)$ denotes the operator of forward propagation as described in the main text by Eq. (4). In the simulation, the following steps are carried out:

(1) $U_1(X,Y) = L(A(x,y))$ is simulated, where $A(x,y)$ is the distribution of the aperture over which graphene is supported. $A(x,y)$ is a round aperture, with transmission equal to 1 inside the aperture and 0 outside the aperture; on the edges of the aperture the transmission smoothly changes from 1 to 0 by applying a cosine-window apodization function[24].

(2) $U_2(X,Y) = L(G(x_i,y_i)) = L(A(x,y)G_0(x_i,y_i,z_i))$ is simulated, where $G_0(x_i,y_i,z_i)$ is 1 at the position $(x_i,y_i,z_i)$ of carbon atom $i$ and 0 elsewhere. $G(x_i,y_i,z_i) = A(x,y)G_0(x,y,z)$ is the distribution of the carbon atoms within the aperture. In our simulations, the carbon atoms in graphene are represented by their coordinates, not as pixels. For each carbon atom within the aperture, the scattered wave is simulated at the detector plane. These waves are then added together giving the total scattered wave.

(3) The intensity distribution at the screen is simulated as $I(X,Y) = |U_1(X,Y) - U_2(X,Y)|^2$.

*Simulation of graphene with adsorbate*: In step (1) of the simulation, the adsorbate distribution is included in $A(x,y)$, and in step (2) the carbon atoms that are at the same position as the adsorbate are excluded from the distribution $G(x,y,z)$.

*Simulation of an out-of-plane ripple*: The positions of carbon atoms are shifted along the $z$-direction in the form of a ripple; the distribution of carbon atoms' coordinates $G_0(x,y,z)$ is accordingly modified.

*Simulation of an in-plane ripple:* The positions of carbon atoms are shifted laterally in the $(x, y)$-plane in the form of a ripple; and the distribution $G_0(x, y, z)$ is accordingly modified.

**REFERENCES**

1. Mermin, N. D. Crystalline order in two dimensions. *Phys. Rev.* **176**, 250–254 (1968).
2. Fasolino, A., Los, J. H. & Katsnelson, M. I. Intrinsic ripples in graphene. *Nature Mater.* **6**, 858–861 (2007).
3. Meyer, J. C. *et al.* On the roughness of single- and bi-layer graphene membranes. *Solid State Commun.* **143**, 101–109 (2007).
4. Meyer, J. C. *et al.* The structure of suspended graphene sheets. *Nature* **446**, 60-63 (2007).
5. Binnig, G. & Rohrer, H. Scanning tunneling microscopy. *IBM Journal of Research and Development* **30**, 355–369 (1986).
6. Binnig, G., Quate, C. F. & Gerber, C. Atomic force microscope. *Phys. Rev. Lett.* **56**, 930–933 (1986).
7. Geringer, V. *et al.* Intrinsic and extrinsic corrugation of monolayer graphene deposited on SiO2. *Phys. Rev. Lett.* **102**, 076102 (2009).
8. Zan, R. *et al.* Scanning tunnelling microscopy of suspended graphene. *Nanoscale* **4**, 3065–3068 (2012).
9. Breitwieser, R. *et al.* Flipping nanoscale ripples of free-standing graphene using a scanning tunneling microscope tip. *Carbon* **77**, 236–243 (2014).
10. Smolyanitsky, A. & Tewary, V. K. Manipulation of graphene's dynamic ripples by local harmonic out-of-plane excitation. *Nanotechnology* **24**, 055701 (2013).
11. Xu, P. *et al.* Unusual ultra-low-frequency fluctuations in freestanding graphene. *Nat. Commun.* **5**, 3720 (2014).
12. Germann, M., Latychevskaia, T., Escher, C. & Fink, H.-W. Nondestructive imaging of individual biomolecules. *Phys. Rev. Lett.* **104**, 095501 (2010).
13. Gabor, D. A new microscopic principle. *Nature* **161**, 777–778 (1948).
14. Gabor, D. Microscopy by reconstructed wave-fronts. *Proc. R. Soc. A* **197**, 454–487 (1949).
15. Stocker, W., Fink, H.-W. & Morin, R. Low-energy electron and ion projection microscopy. *Ultramicroscopy* **31**, 379–384 (1989).
16. Stocker, W., Fink, H.-W. & Morin, R. Low-energy electron projection microscopy. *Journal De Physique* **50**, C8519–C8521 (1989).
17. Fink, H.-W., Stocker, W. & Schmid, H. Holography with low-energy electrons. *Phys. Rev. Lett.* **65**, 1204–1206 (1990).

18  Chang, W.-T. *et al.* Low-voltage coherent electron imaging based on a single-atom electron source. *arXiv*, 1512.08371 (2015).

19  Fink, H.-W. Mono-atomic tips for scanning tunneling microscopy. *IBM Journal of Research and Development* **30**, 460–465 (1986).

20  Fink, H.-W. Point-source for ions and electrons. *Physica Scripta* **38**, 260–263 (1988).

21  Kuo, H. S. *et al.* Preparation and characterization of single-atom tips. *Nano Lett.* **4**, 2379–2382 (2004).

22  Chang, C. C., Kuo, H. S., Hwang, I. S. & Tsong, T. T. A fully coherent electron beam from a noble-metal covered W(111) single-atom emitter. *Nanotechnology* **20**, 115401 (2009).

23  Latychevskaia, T., Wicki, F., Longchamp, J.-N., Escher, C. & Fink, H.-W. Direct observation of individual charges and their dynamics on graphene by low-energy electron holography. *Nano Lett.* **16**, 5469–5474 (2016).

24  Latychevskaia, T. & Fink, H.-W. Practical algorithms for simulation and reconstruction of digital in-line holograms. *Appl. Optics* **54**, 2424–2434 (2015).

25  Latychevskaia, T. & Fink, H.-W. Reconstruction of purely absorbing, absorbing and phase-shifting, and strong phase-shifting objects from their single-shot in-line holograms. *Appl. Optics* **54**, 3925–3932 (2015).

**ACKNOWLEDGEMENTS**

This research is supported by Academia Sinica of R. O. C. (AS-102-TP-A01).

**AUTHOR CONTRIBUTION**

IS.H. initiated the project of low-energy electron imaging; WH.H., WT.C. and CY.L. designed and performed experiments; WH.H. acquired the DBED patterns; IS.H. contributed to writing the manuscript; WH.H., WT.C., CY.L., and IS.H. participated in discussions. T.L. proposed explanation of the observed experimental effects, developed the theory, made the simulations, analysed the experimental data and wrote the manuscript.

# SUPPLEMENTARY

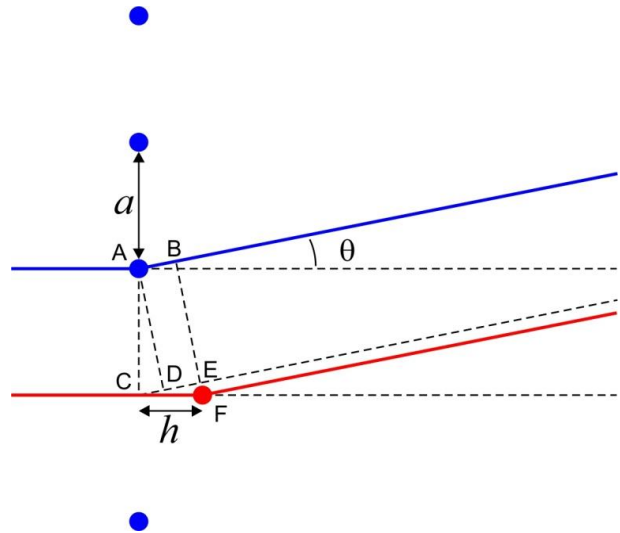

**Supplementary Figure 1. Phase change at an out-of-plane ripple.** We consider scattering off two atoms at different *z*-positions in an out-of-plane ripple, atoms A and F. The difference in the optical path length between the wave scattered off atom A and atom F equals: $\Delta s = \text{CF} - \text{AB}$, where $\text{AB} = \text{CE} - \text{CD} = h\cos\vartheta - a\sin\vartheta$, so that $\Delta s = h - h\cos\vartheta + a\sin\vartheta = h(1-\cos\vartheta) + a\sin\vartheta$, and the phase shift equals $\Delta\varphi = \frac{2\pi}{\lambda}\Delta s = \frac{2\pi}{\lambda}h(1-\cos\vartheta) + \frac{2\pi}{\lambda}a\sin\vartheta$. Without a ripple, the phase shift is given by $\Delta\varphi = \frac{2\pi}{\lambda}a\sin\vartheta$, which is the phase shift gained by conventional diffraction on periodical lattice. When a ripple is present $h \neq 0$ and the phase shift has additional term $\frac{2\pi}{\lambda}h(1-\cos\vartheta)$, which is pronounced at higher scattering angles.

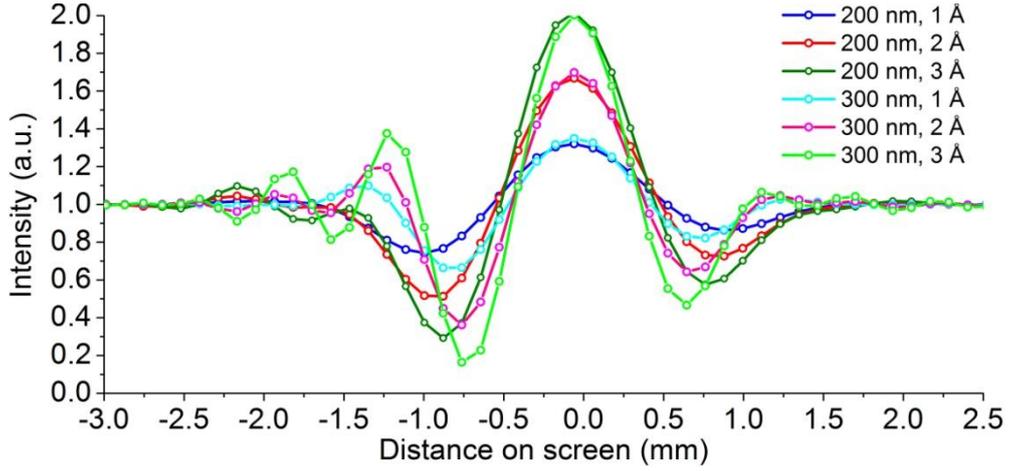

**Supplementary Figure 2. Intensity profiles through the centre of the (-1010) divergent beam electron diffraction (DBED) spot at different simulation parameters.** Ripple amplitude is 1, 2 and 3 Å, and the source-to-sample distance is 200 nm and 300 nm. In these simulations, the source-to-detector distance is 50 mm, and the electron energy is 230 eV. It can be seen that the ripple with higher amplitude results in an increased intensity contrast, which can be intuitively expected. Also, the same ripple can produce different contrast at different source-to-sample distances: at larger source-to-sample distance, the width of the intensity distribution in the first-order diffraction spot is de-magnified due to the decreased magnification, but the contrast becomes higher.

**Supplementary Note 1**

**Imaging weak phase objects**

The transmission function of a weak phase object can be approximated as:

$$t(x,y) = e^{i\Delta\varphi(x,y)} \approx 1 + i\Delta\varphi(x,y), \tag{1}$$

where $(x, y)$ are the coordinates in the object plane and $\Delta\varphi(x, y)$ is the phase shift introduced to the incident wave. Since a constant phase shift can be added to a wave without changing its intensity distribution, the phase distribution superimposed onto the incident wave $\Delta\varphi(x, y)$ can be re-written so that $\Delta\varphi(x, y) > 0$. Equation (1) allows for the splitting of the exit wave into two terms: the number 1 describes the reference wave and $i\Delta\varphi(x, y)$ describes the perturbation to the reference wave caused by the object and thus can be interpreted as the object wave. Because the phase change superimposed onto the incident wave in the object domain is preserved while the wave propagates toward the detector we can write

$$L(1 + i\Delta\varphi(x,y)) \rightarrow 1 + i\Delta\varphi(X,Y), \tag{2}$$

where $L$ is an operator of forward propagation towards the detector plane, and $(X,Y)$ are the coordinates in the detector plane. The intensity on the detector is given by:

$$I(X,Y) = |1 + i\Delta\varphi(X,Y)|^2 = 1 + |\Delta\varphi(X,Y)|^2. \tag{3}$$

Equation (3) describes how the three-dimensional surface of a ripple is transformed into the contrast of the intensity distribution. For a ripple with low amplitude $h$, the approximation of weak phase shift applies and Eq. (3) can predict the intensity contrast caused by the ripple:

$$\Delta I(X,Y) = |\Delta\varphi(X,Y)|^2 \approx \left(\frac{2\pi}{\lambda} h(1-\cos\vartheta)\right)^2.$$